\documentclass[pdflatex,sn-mathphys]{sn-jnl}

\usepackage{xspace}

\jyear{2023}%

\theoremstyle{thmstyleone}%
\theoremstyle{thmstyletwo}%

\def\TeV{\ensuremath{\mathrm{TeV}\xspace}}
\def\GeV{\ensuremath{\mathrm{GeV}\xspace}}

\newcommand\vkin[3]{\ensuremath{#1_\mathrm{#2}^{#3}\xspace}}

\newcommand\xrec[1]{\vkin{x}{reco}{#1}}
\newcommand\yrec[1]{\vkin{y}{reco}{#1}}
\newcommand\Erec[1]{\vkin{E}{reco}{#1}}
\newcommand\xloc[1]{\vkin{x}{loc}{#1}}
\newcommand\yloc[1]{\vkin{y}{loc}{#1}}
\newcommand\xwin[1]{\vkin{x}{win}{#1}}
\newcommand\ywin[1]{\vkin{y}{win}{#1}}
\newcommand\xgen[1]{\vkin{x}{gen}{#1}}
\newcommand\ygen[1]{\vkin{y}{gen}{#1}}
\newcommand\Egen[1]{\vkin{E}{gen}{#1}}

\theoremstyle{thmstylethree}%

\begin{document}

\title[Reconstruction of electromagnetic showers in calorimeters using Deep Learning]{Reconstruction of electromagnetic showers in calorimeters using Deep Learning}


\author[1]{\fnm{Polina} \sur{Simkina}}\email{simkina.polina@gmail.com}

\author[1]{\fnm{Fabrice} \sur{Couderc}}\email{fabrice.couderc@cea.fr}

\author[1]{\fnm{Julie} \sur{Malcl\`es}}\email{julie.malcles@cea.fr}

\author[1]{\fnm{Mehmet \"Ozg\"ur} \sur{Sahin}}\email{ozgur.sahin@cea.fr}

\affil[1]{\orgdiv{IRFU}, \orgname{CEA, Université Paris-Saclay}, \orgaddress{\city{Gif-sur-Yvette}, \postcode{F-91191}, \country{France}}}


\abstract{The precise reconstruction of properties of photons and electrons in modern high energy physics detectors, such as the CMS or Atlas experiments, plays a crucial role in numerous physics results. Conventional geometrical algorithms are used to reconstruct the energy and position of these particles from the showers they induce in the electromagnetic calorimeter. Despite their accuracy and efficiency, these methods still suffer from several limitations, such as low-energy background and limited capacity to reconstruct close-by particles. This paper introduces an innovative machine-learning technique to measure the energy and position of photons and electrons based on convolutional and graph neural networks, taking the geometry of the CMS electromagnetic calorimeter as an example. The developed network demonstrates a significant improvement in resolution both for photon energy and position predictions compared to the algorithm used in CMS. Notably, one of the main advantages of this new approach is its ability to better distinguish between multiple close-by electromagnetic showers.}

\keywords{electromagnetic calorimeter, energy clustering, machine learning, convolutional neural networks, graph neural networks}



\maketitle

\section{Introduction}\label{sec1}

In modern hadron collider experiments with a large number of simultaneous collisions and a large number of detector cells, reconstructing the properties of individual particles from the information collected by the detectors poses a significant computational challenge. These experiments use, among other sub-detectors, electromagnetic calorimeters that record the energy deposits left by particles and measure in particular the energy of electrons and photons. Generally, energy deposits in different cells of these detectors must be clustered together to reconstruct the energy and position of the initial particle. Such reconstruction is a complex problem due to the high multiplicity of particles and the overlap of showers, which can be further investigated by taking the Electromagnetic CALorimeter (ECAL) of the Compact Muon Solenoid (CMS) detector as an example.


The CMS experiment is a general-purpose detector situated at the Large Hadron Collider (LHC)~\cite{cms} at CERN. Its objectives are to probe the main theoretical framework used in particle physics (the Standard Model) and search for physics beyond it. To do so, it is necessary to detect, identify, and determine the kinematic properties of particles produced by proton-proton collisions at center-of-mass energies reaching up to $13.6~\TeV$. 

Numerous physics analyses performed with the data collected by the CMS detector (\emph{e.g.} relating to the study of the properties of the Higgs boson~\cite{higgs},~\cite{higgs_mass}) require precise reconstruction of photon and electron properties. It is done in a multi-step process starting from the reconstruction of single energy deposits in the detector and ending with the identification of the particle type and its origin~\cite{CMS:2020uim}. 

The kinematic properties of individual photons are evaluated using a traditional geometrical algorithm that clusters the energies they deposit in different cells of the detector. The algorithm used by CMS is described in Section~\ref{section2} and is called \textit{PFClustering}, \textit{PF} standing for \textit{Particle-Flow}~\cite{pfclustering}. While this method is accurate and efficient, it has limitations: 

\begin{enumerate}
    \item It has a limited ability to accurately distinguish two close-by photons, such as photons produced in neutral pion ($\pi_{0}$) decays, which can mimic the energy pattern of an isolated $\gamma$, or decays of potential new particles, for instance, the exotic decay of the Higgs boson $H \rightarrow AA \rightarrow 4\gamma$~\cite{exotic_higgs}, where $A$ is a light scalar or pseudoscalar particle. 
    \item The algorithm has a high background rate at very low energies ($< 1$~GeV) related to experimental noise. Such backgrounds are expected to increase with the aging of the detector and the performance is expected to further deteriorate. 
\end{enumerate}

In order to overcome these limitations and improve the reconstruction of particle properties, machine learning (ML) algorithms can be considered. They have been already widely applied in physics analyses based on photons and electrons in CMS (e.g. \cite{art1}, \cite{art2}).
However, ML techniques have not been used in CMS for energy clustering, nor optimized for multiple overlapping energy deposits originating from several particles, where new challenges emerge for the model performance and from computer-related constraints.

In this paper, we present a novel ML algorithm, named DeepCluster, to reconstruct energy deposits in electromagnetic calorimeters. This algorithm is based on convolutional and graph neural networks. The developed model significantly outperforms the PFClustering algorithm in terms of energy and position resolution, while also mitigating the limitations of the traditional approach regarding close-by particle identification. We show the different steps required to develop the final algorithm, explaining the reasoning behind the choice of specific methods. Finally, the performance is shown for various particle types (photons, electrons, and neutral pions), and compared to the performance of the PFClustering algorithm. 

\section{Particle reconstruction in the CMS electromagnetic calorimeter} \label{section2}


 The method described in this paper has broad applicability across a variety of clustering tasks. The CMS ECAL serves as a critical instrument for the precise detection and identification of electrons and photons, thus acting as the reference case for the proposed methodology. We initially provide an overview of this detector's operational principles and geometric configuration. Then, we delve into the PFClustering algorithm, which is currently employed by the CMS collaboration to reconstruct the properties of electromagnetic deposits. Finally, we discuss the utilized energy correction techniques. This comprehensive overview serves as a foundation for comparison with the novel methodology proposed in this study.

\subsection{CMS ECAL}
The CMS ECAL is a homogeneous calorimeter consisting of about 75,000 lead tungstate ($\mathrm{PbWO}_{4}$) scintillating crystals. 
It is divided into two main parts: the barrel (crystal size: 2.2 $\times$ 2.2 $\times$ 23~$\mathrm{cm}^3$), covering the pseudorapidity region $\lvert \eta \rvert$ $<$ 1.479, and the endcaps (crystal size: 2.9 $\times$ 2.9 $\times$ 23~$\mathrm{cm}^3$), covering the pseudorapidity regions 1.479 $<$ $\lvert \eta \rvert$ $<$ 3.0. In the barrel, the crystal length corresponds to 25.8 radiation length $X_0$ and the crystal transverse size matches the small Moli\`ere radius of PbWO4. A complete description of the CMS experiment and the ECAL is given in Ref.~\cite{cms}. Crystals are arranged in a quasi-projective geometry with regard to the center of the interaction region: crystals are tilted by 3 degrees with respect to this point to avoid leakage in mechanical gaps.

The main purpose of the ECAL is the characterization of electromagnetic energy deposits. When a photon or an electron enters the calorimeter, it starts an electromagnetic shower \cite{calorimetry}, ultimately generating scintillation photons that are detected by photodetectors. This energy shower can be spread among multiple crystals around the entry point. A combination of adjacent-crystal energy deposits is called a cluster. The goal of reconstruction algorithms is to estimate as precisely as possible the energy and position of the primary particle entering the calorimeter from the corresponding cluster properties.

\subsection{The PFClustering algorithm}
The PFClustering algorithm~\cite{pfclustering} is designed to ensure high efficiency even for low-energy particles. The energy clusters reconstructed with the PFClustering algorithm are called PFClusters and they are formed from the following steps:
\begin{enumerate} 
    \item Energy deposits in each crystal left by particles entering the detector are reconstructed as \textit{hits}.
    \item {Hits} with energy exceeding a certain threshold ($E_\mathrm{thr}^\mathrm{seed}$) and larger than the energy of adjacent {hits} (either sharing a side or a corner) are selected as \textit{seeds}. 
    \item Each seed is combined with its eight neighboring {hits} to create a \textit{topological cluster}.
    \item Topological clusters are grown by aggregating all {hits} with at least a corner or a side in common with a cell already in a cluster. For a crystal to be included in a topological cluster, its energy must exceed another specified energy threshold ($E_\mathrm{thr}^\mathrm{gather}$).
    \item One topological cluster may contain multiple {seeds} (each {seed} potentially corresponding to a single particle), and in this case, energy deposits originating from distinct particles could overlap within different {hits}. A PFCluster is attributed to each seed. The fraction $f_{ji}$ of energy of each {hit} $j$ attributed to PFCluster $i$ in the topological cluster is determined from an expectation-minimization iterative algorithm rooted in a Gaussian-mixture model. Its detailed description can be found in Ref.~\cite{pfclustering}. 
\end{enumerate}

The position ($\vec{\mu}_{\mathrm{reco}}^i$) and predicted energy ($\Erec{i}$) of the $i^{\mathrm{th}}$ cluster are evaluated following the formulas:

    \begin{equation}
E_{\mathrm{reco}}^i=\sum_{j=1}^M f_{j i} A_j
\label{gaussian_energy}
\end{equation}

    \begin{equation}
\vec{\mu}_{\mathrm{reco}}^i=\frac{1}{E_{\mathrm{reco}}^i}\sum_{j=1}^M f_{j i} A_j \vec{c}_j, 
\label{gaussian_coordinate}
\end{equation}
where $M$ is the total number of {hits} in a topological cluster, $A_j$ is the deposited energy in $j_{\mathrm{th}}$ {hit}, $\vec{c_{j}}$ is the position of the $j_{\mathrm{th}}$ {hit}.



The parameters of the PFClustering algorithm used in this paper correspond to the tuning performed for the Run~3 (2022-2025 operation) of the LHC~\cite{clusterin_run3}: $E_\mathrm{thr}^\mathrm{seed}=3~\sigma_n$,
$E_\mathrm{thr}^\mathrm{gather}=0~\GeV$, with $\sigma_n$ the noise parameter discussed in Section~\ref{section:dataset}.

\subsection{Energy correction}
Due to the possibility of energy leakage from the shower and the per-crystal energy thresholds implemented in the reconstruction step, the particle energy evaluated by the PFClustering algorithm tends to be underestimated. To compensate for this, an energy correction is applied to the PFClusters obtained in the preceding section. In the CMS experiment, this correction is carried out using a multivariate technique known as the Boosted Decision Trees (BDT)~\cite{BDT} and trained on simulation. A comprehensive list of all inputs to the BDT used in this study with their brief descriptions is presented in Table~\ref{table:bdt_parameters}. 

\begin{table}[ht]
    \centering
    \renewcommand{\arraystretch}{1.5}
    \setlength{\tabcolsep}{10pt}
    \begin{tabular}{l|p{0.52\linewidth}}
    \toprule
    \textbf{Parameter} & \textbf{Description} \\
    \midrule
    $E_{\mathrm{reco}}$ & Energy predicted by the PFClustering algorithm \\
    $x_{\mathrm{reco}}$, $y_{\mathrm{reco}}$ & Position predicted by the PFClustering algorithm \\
    $E_{\mathrm{max}}$, $E_{2\mathrm{nd}}$ & Largest and second largest energy deposits in the crystals within the processed PFCluster. \\
    $E_{\mathrm{LR}}$ & The energy deposit difference between the left and right crystals in relation to the seed crystal. \\
    $E_{\mathrm{TB}}$ & The energy deposit difference between the top and bottom crystals in relation to the seed crystal. \\
    $cov(X, X)$, $cov(X, Y)$, $cov(Y, Y)$ & Covariance values between the PFCluster spread in different directions.  \\
    $E_{3\times3}$ & The sum of energies in a 3$\times$3 matrix around the seed crystal. \\
    \bottomrule
    \end{tabular}
    \caption{Input variables to the PFClustering energy regression BDT.}
    \label{table:bdt_parameters}
\end{table}

While the reconstruction algorithm implemented in the CMS experiment utilizes a limited subset of variables for energy correction, we have chosen to incorporate additional variables to enhance its performance.
Detailed information on the training and optimization of the BDT is available in \nameref{appendix_a}.
The correction is applied when comparing the energy resolution of the PFClustering algorithm to the one of the DeepCluster algorithm in Section \ref{section:results}.

\section{Simulation and dataset}\label{section:dataset}
To train the DeepCluster model and compare its performance to the ones of the PFClustering algorithm, a calorimeter simulation, referred to as a \textit{toy calorimeter}, is created using the Geant4 software \citep{geant}.

This simulation uses a simple geometry made of a rectangular shape consisting of $51 \times 51$ crystals, preserving the physical characteristics of the barrel part of ECAL: PbWO$_4$ crystals with dimensions of $2.2 \times 2.2 \times 23$~cm$^3$. In each crystal, the measured energy is derived by randomly smearing the deposited energy as obtained from Geant4 (see section~\ref{sec:energy_resolution}). In the simulation, we neglect the ECAL-crystal tilt. Furthermore, additional interactions (pile up) are not considered. 

We first create an initial dataset composed of single high-energy $\gamma$ directed perpendicularly towards the detector surface.
Their energies are uniformly distributed within the range of $1$-$100~\GeV$.
The position at which the particles enter the toy calorimeter is randomly chosen, avoiding the edges of the detector to ensure that the particle deposits most of its energy in the active material of the calorimeter.

We demonstrate the validity of our toy detector by comparing the energy deposit profiles obtained with our simulation to the ones from electron-test-beam data, see~\nameref{appendix_b}.

\subsection{Datasets}
In order to train and test the DeepCluster model,
we create two separate datasets (using pandas~\cite{pandas} and numpy~\cite{numpy} libraries from Python programming language) from the aforementioned initial dataset: 
\begin{itemize}
    \item \textbf{Single-photon dataset}. \newline Each entry of the dataset consists of one photon. It is used both for training and as a primary check of DeepCluster performance regarding coordinate and energy resolution. It corresponds to the case of isolated particles in the calorimeter.
    \item \textbf{Two-photon dataset}. \newline Each entry is created by superimposing two different samples from the initial dataset. In this dataset, only particles with positions located within a maximum distance ($\Delta R = \sqrt{\Delta x^2 + \Delta y^2}$) of 3 crystals from each other are selected, excluding cases where the two particles enter the calorimeter in the same crystal.  This dataset represents two close-by photons in the calorimeter, mimicking the signature of a $\pi_0$ decay, for instance. 
\end{itemize} 

In this study, the training (resp. validation) dataset consists of a random mixture of 600k (resp. 200k) samples from the single-photon dataset and 300k (resp. 100k) samples from the two-photon dataset. The training dataset is used to train the DeepCluster model while the validation dataset is used to select the best model ensuring no overtraining. In addition, we estimate the performance with a test dataset composed of 100k single photons and 50k two-photon samples.

Moreover, in order to gain a comprehensive understanding of the model performance across various potential use cases, extra datasets used only for evaluation are created: 
\begin{itemize}
     \item \textbf{Electron} dataset. \newline The sample is created in a similar way as the photon sample. Each electron has an energy randomly chosen from a uniform distribution in the range $[1, 100]~\GeV$. Since electrons are primarily reconstructed within the ECAL, it is important to check the algorithm's performance for these particles. 
   \item \textbf{Neutral pion} dataset.  \newline To create this dataset, we simulate $\pi^{0}$s emitted at a distance of $130~\mathrm{cm}$ from the toy calorimeter front face and with energies uniformly distributed in the range $[1, 100]~\GeV$. It consists of 180k samples.
\end{itemize}


\subsection{Per-crystal energy}
\label{sec:energy_resolution}

Finally, as the simulation does not include all the steps of the ECAL readout chain, the per-crystal energy is obtained by smearing the true deposited energy. This provides a realistic simulation of the calorimeter response and energy resolution, including the simulation of the readout-electronic noise. 

The ECAL energy resolution is parametrized as follows \cite{cms_tdr}: 
\begin{equation}
    \left(\frac{\sigma_E}{E}\right)^2 = \left(\frac{a}{\sqrt{E}}\right)^2 + \left(\frac{\sigma_n}{E}\right)^2 + c^2, 
\label{eq:energy_resolution}
\end{equation}

where $a$, $\sigma_n$, and $c$ are called respectively stochastic, noise, and constant terms. The energy $E_\mathrm{xtal}$ measured in a crystal $xtal$ is given by: 
\begin{equation}
    E_\mathrm{xtal} = E_\mathrm{xtal}^\mathrm{true} \times \mathcal{N}\left(\mu=1; \sigma=\frac{\sigma_E}{E}\right),
\end{equation}
where $\mathcal{N}(\mu; \sigma)$ is a random number following a Gaussian distribution with mean $\mu$ and standard deviation $\sigma$, and $E_\mathrm{xtal}^\mathrm{true}$ is the true energy deposited in the crystal $xtal$ obtained from the Geant4 program. The chosen parameters correspond to the ECAL conditions for Run~3~\cite{ecal_constant, clusterin_run3}: $a$~=~0.03~GeV$^{-\frac{1}{2}}$, $\sigma_n$ = 0.167~GeV, $c$ = 0.0035. A cut at 50~MeV is applied on the smeared energy to mitigate the noise.

\section{DeepCluster model}
There has been a surge in the application of machine learning techniques in the field of particle physics over recent years due to the capacity of these algorithms to extract complex patterns. 
In particular, techniques such as deep neural networks (DNN) and convolutional neural networks (CNN) have enabled unprecedented levels of accuracy and efficiency in tasks such as particle tracking and calorimetry clustering. Graph neural networks (GNN) have also recently gained significant popularity in high-energy physics applications (e.g.~\cite{graph_particle},~\cite{graph_2}) due to the following factors:
\begin{itemize}
    \item they can handle sparse data coming from complex detector geometries,
    \item they are applicable to non-Euclidean data structures with variable input sizes.
\end{itemize}

To achieve optimal performance, we combine multiple state-of-the-art machine learning methods in the DeepCluster architecture.
First, we use the excellent pattern recognition abilities of CNNs. 
As energy deposited in the crystals of the calorimeter can be represented as pixel intensities of an image, CNN can be easily and naturally applied to treat calorimeter data. 
Secondly, we use a GNN to allow information transfer between neighboring particles.
\\


In the DeepCluster model, the particle-reconstruction task is divided into two consecutive steps: 
\begin{enumerate}
     \item Extract small windows ($7\times7$ crystals), named \textit{seed windows}, whose energy deposits potentially originate from a real particle and not from noise. This is performed by a first NN called the \textit{seed-finder NN} described in Section~\ref{sec:seed_finder}.
     \item For each seed window predict the kinematic properties of the corresponding generated particle. This is done with a second NN called \textit{center-finder NN}. In this work, we develop two different approaches for the center-finder NN:
     \begin{itemize}
     \item The first one, based on a CNN, is described in Section~\ref{sec:center_CNN}. 
     \item To circumvent the limitations of this CNN-based center-finder, a second center-finder, using a GNN, is introduced in Section~\ref{section:cf_gnn}.
     \end{itemize}
 \end{enumerate}
With this approach, the networks process only small crystal matrices as opposed to the full calorimeter matrix ($51\times51$ crystals). This significantly reduces the need for computational power and allows to easily scale this method to a real detector (the ECAL barrel contains $170\times360$ crystals). 

\subsection{\textit{Seed windows} definition}\label{section:clustering_truth_association}

\begin{wrapfigure}[14]{l}{0.5\textwidth}
  \begin{center}
    \includegraphics[width=0.48\textwidth]{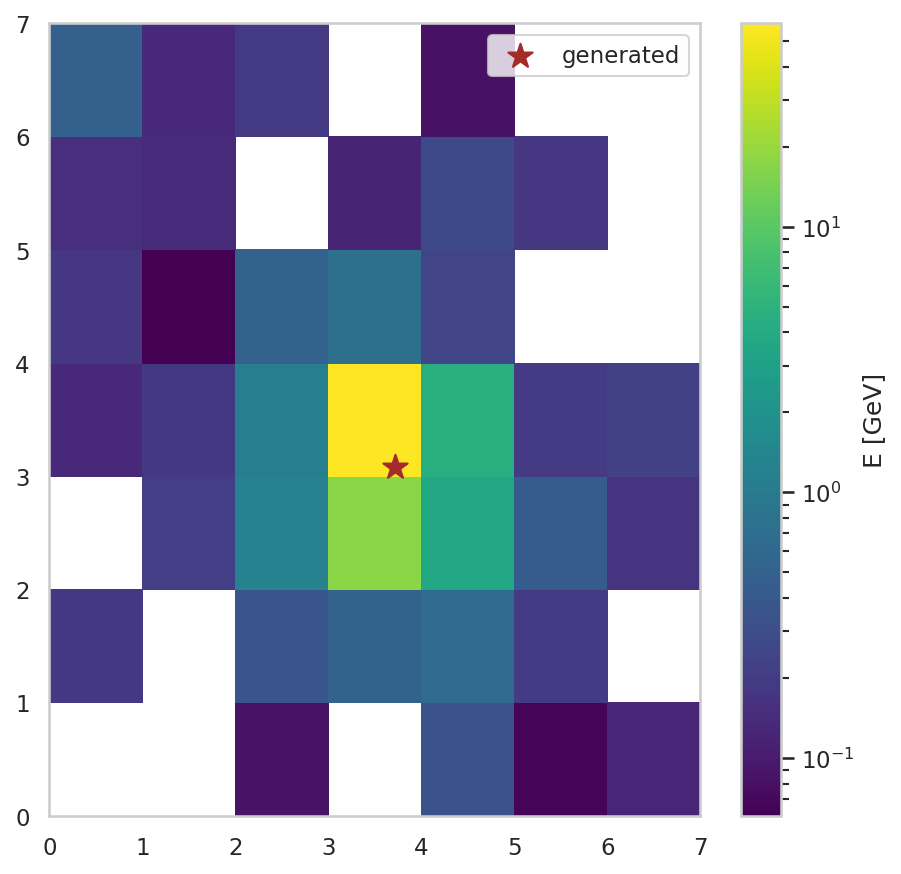}
  \end{center}
  \caption{Example of a seed window.}
  \label{fig:event_display_7}
\end{wrapfigure}

The inputs to the DeepCluster model are called \textit{seed windows} and are obtained for each sample as follows:
\begin{enumerate}
    \item All the crystals from the toy calorimeter with energy deposits $E_{xtal}>0.5~\GeV$ are selected and defined as \textit{seed crystals}. The selected $0.5~\GeV$ threshold value corresponds to $E_\mathrm{thr}^\mathrm{seed}$ used in the PFClustering algorithm and tuned for Run~3 operations. 
    \item For each seed crystal, a seed window is created. This is a matrix of size $7\times7$ crystals centered on the seed crystal. 
    An example of a seed window is shown in Fig.~\ref{fig:event_display_7}.
\end{enumerate}

From one simulated sample corresponding to the full toy calorimeter, several seed windows can be created. They originate from a real particle or from noise.

In order to train the different networks, we need to associate  the seed windows to their corresponding generated particles and the subsequent truth labeling has to be defined. We first  check if the impact position of any generated particle in the considered sample lies within the boundaries of the seed crystal:
\begin{itemize}
\item In such a case, the corresponding particle (at most one by construction) is associated to the seed window. The window is labeled \textit{true seed window} and assigned three kinematic variables corresponding to the true particle: generated position $(\xgen{}, \ygen{})$ and energy \Egen{}.
\item Otherwise, it is labeled as background. 
\end{itemize}
For {true seed windows}, the local position of the generated particle inside the seed window (\xloc{}, \yloc{}) is defined as:
\begin{equation}
\begin{split}
\xloc{} &= \xgen{} - \xwin{}\\
\yloc{} &= \ygen{} - \ywin{}\\
\end{split}
\label{eq:x_center}
\end{equation}
where (\xwin{}, \ywin{}) corresponds to the position of the seed window center.

\subsection{Seed-finder NN}
\label{sec:seed_finder}
The seed-finder NN is the first network in the DeepCluster model. This is a CNN, whose goal is to select the true seed windows and discard background ones. 

The network takes a seed window as input and assigns to it a seed-finder score ($P_\mathrm{seedSF}$), indicating the likelihood to be a true seed window. The seed windows with $P_\mathrm{seedSF} < P^\mathrm{thr}_\mathrm{seedSF}$, where $P^\mathrm{thr}_\mathrm{seedSF}$ is a tunable threshold defined in Section~\ref{section:optimization}, are discarded, the other ones are passed to the center-finder NN.

\begin{figure}[ht]
\centering
\includegraphics[width=0.85\linewidth]{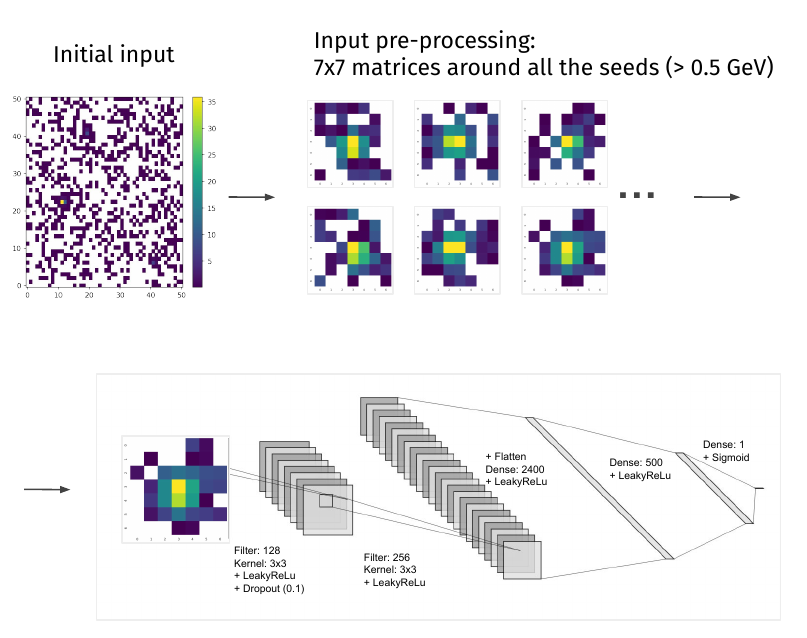} 
\caption{Seed-finder NN architecture. $7\times7$ seed windows are first selected around all possible seeds ($E_{xtal}>0.5~\GeV$). They are separately passed as input to the seed-finder NN. The input is processed by two convolutional layers until the vector of summary features is extracted. This vector is further passed to two dense layers, resulting in the network output: the seed-finder score $P_\mathrm{seedSF}$. It represents the likelihood of the input to originate from a generated particle. Detailed information on the number of nodes at each layer is presented in the figure.}
\label{figure:seed_finder_architecture}
\end{figure}

The seed-finder-NN architecture consists of two convolutional and two dense layers. LeakyReLU is chosen as the activation function~\cite{relu} and a dropout of 10$\%$ is applied after the first convolutions. For the output of the seed-finder NN, a sigmoid activation function is used. A detailed view of the model architecture is presented in Fig.~\ref{figure:seed_finder_architecture}. 

The model is trained using the Adam optimizer \cite{adam} with a learning rate of 0.0001 and a batch size of 64. We use the binary cross entropy as a loss function. The network is trained for $\approx$100 epochs and the epoch yielding the best result on the validation sample is chosen.

Compared to the seeding step used in the PFClustering algorithm, the seed-finder NN provides several advantages:
\begin{itemize}
    \item As the condition for the seed to be a local maximum is removed, the seed-finder NN provides a better possibility to reconstruct close-by photons.  
    \item The seed-finder NN performs a refined seed window selection that helps to significantly eliminate the low-energy background coming from electronic noise. 
\end{itemize}

\subsection{Center-finder NN -- Convolutional Neural Network}
\label{sec:center_CNN}

The center-finder NN is the second step of the DeepCluster model. It predicts the position (\xrec{i}, \yrec{i}) and energy \Erec{i} of the generated particle associated to the seed window $i$; the corresponding generated quantities are named (\xloc{i}, \yloc{i}), \Egen{i}. The global coordinates can be further inferred by inverting Eq.~\ref{eq:x_center}.

Similarly to the seed-finder NN, the CNN-based center-finder takes seed windows as input. All the inputs are processed independently from each other. In the training phase, the inputs are limited to the true seed windows. However, for the evaluation, the center-finder NN processes all the seed windows with the seed-finder scores passing $P_\mathrm{seed}^\mathrm{thr}$. 

\begin{figure}
\centering
\includegraphics[width=\linewidth]{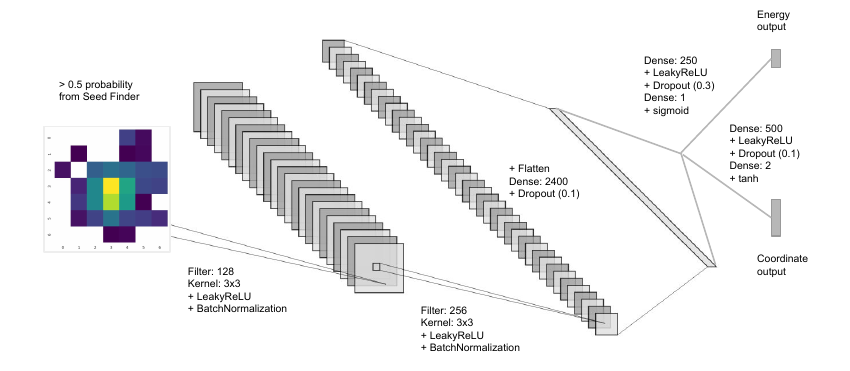} 
\caption{Center-finder NN architecture. The seed windows are passed separately as input to the network. Each input is processed by two convolutional layers until the vector of summary features is extracted. This vector is passed through one dense layer and further sent separately to two different branches (coordinate and energy predictions). In each branch, it passes through two additional dense layers. Detailed information on the number of nodes at each layer is presented in the figure.}
\label{figure:center_finder_architecture}
\end{figure}

The architecture of the center-finder NN is close to the one of the seed-finder NN: it consists of multiple convolutional layers, followed by dense layers that are divided into two parts: one resulting in coordinate prediction and another in energy prediction. ReLU activation function is applied everywhere except for the output layer, where $\tanh$ and sigmoid functions are used respectively for position and energy predictions~\cite{relu}. The dropout level is set to 10$\%$ everywhere except for the last energy prediction layer, where it is set to 30$\%$. The full network architecture with precise details on the number of nodes is presented in Fig.~\ref{figure:center_finder_architecture}. 

The network is trained for 1000 epochs with a batch size $N_b$ of 64 seed windows. We use the mean absolute error loss defined by:
\begin{equation}
\mathcal{L}_{\mathrm{kin}} = \frac{1}{N_b} \sum_{i=1}^{N_b} \frac{1}{2}
\left( \lvert \xrec{i} - \xloc{i} \rvert +  \lvert \yrec{i}-\yloc{i}\rvert \right)
+
\lvert \Erec{i} - \Egen{i}\rvert
\end{equation}
The training is performed using the Adam optimizer with a learning rate of 0.0001. The chosen epoch is the one providing the best performance according to the validation dataset.

\subsubsection{Results}

The results of the DeepCluster model and the PFClustering algorithm are compared with the single- and two-photon test datasets. For the seed-finder NN, we set $P^\mathrm{thr}_\mathrm{seedSF} = 0.3$. This latter threshold is tuned for the final DeepCluster model as discussed in Section~\ref{section:optimization}.

\begin{figure}[ht]
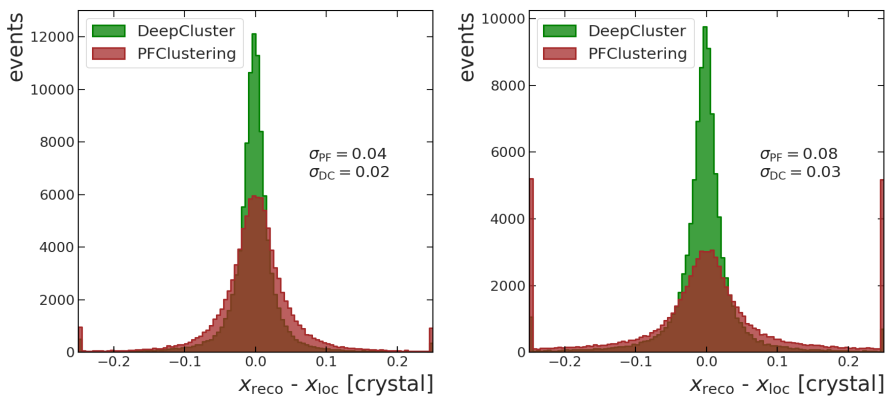

\centering
\includegraphics[width=0.49\linewidth]{figures/x_resolution_cnn_paper_1.pdf} 
\includegraphics[width=0.49\linewidth]{figures/x_resolution_cnn_paper_2.pdf}
\caption{Distribution of the variable $\xrec{}-\xloc{}$. The results are obtained by applying the PFClustering algorithm and DeepCluster network on the single-photon (left) and two-photon (right) test datasets. The resolutions (see text) are reported in the figures.}
\label{fig:two_step_coordinate_resolution}
\end{figure}

The performance for the position prediction is presented in Fig.~\ref{fig:two_step_coordinate_resolution} for the single-photon dataset on the left and for the two-photon dataset on the right. Each plot shows the distribution of the difference $\xrec{}-\xloc{}$. Similar results are obtained for the $y$-coordinate. We associate reconstructed objects to generated objects using a matching procedure described in \nameref{appendix_c}.

The DeepCluster network significantly outperforms the PFClustering algorithm. The coordinate resolution (evaluated as half the interval containing 68\% of the distribution and centered on the median)
for the DeepCluster is 0.02~crystal compared to 0.04~crystal for the PFClustering algorithm for the single-photon dataset and 0.03~crystal versus 0.08~crystal for the two-photon dataset. 

The position reconstruction for the two-photon dataset is a more difficult task than for the single-photon one because of overlapping energy clusters for close-by generated particles. This explains the performance degradation in the two-photon dataset.

Concerning the energy prediction, the performance of the PFClustering and DeepCluster algorithms is  closer. They are presented for the optimized DeepCluster algorithm in Section~\ref{section:results}.

\subsubsection{Per-event energy overestimation}
Because the network processes only seed windows, this approach can be easily extended to a real calorimeter. However, it also raises a major issue described in this section. Contrary to the PFClustering, the local maximum condition is omitted for the seed crystal in the DeepCluster. As a consequence, two (or more) neighboring crystals can be selected as seeds. While this allows for efficient reconstruction of close-by particles, this can also create two or more separate seed windows corresponding to a single generated particle, which would share energy among several crystals. For the majority of these latter cases, the seed-finder NN is able to identify the local maximum itself and predict a high seed-finder score $P_\mathrm{seedSF}$ for the corresponding seed window (in which the local maximum is the central crystal) and a low score for its neighbor window.

However, when the position of the generated particle is close to the edge of a crystal, its energy deposit can be very similar in the two neighboring crystals. Both of them get a high $P_\mathrm{seedSF}$, giving rise to two distinct seed windows. These windows are separately passed to the center-finder NN that predicts similar coordinates and energies for both of them. The same generated particle can therefore be reconstructed twice by the DeepCluster algorithm, thus overestimating the total energy reconstructed in the event. In the following, this is referred to as energy \textit{double counting}.

\begin{figure}
\centering
  \begin{minipage}{0.59\textwidth}
    \includegraphics[width=\linewidth]{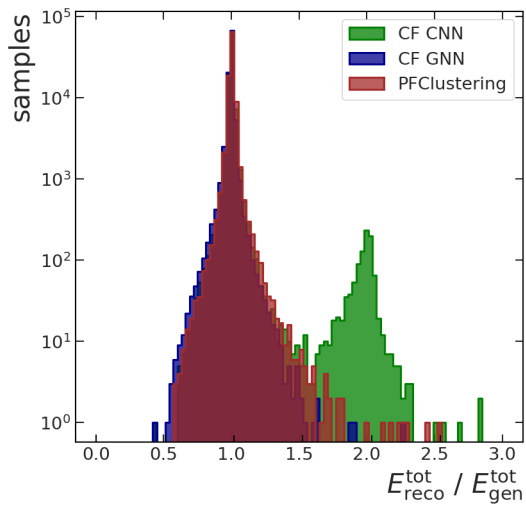} 
  \end{minipage}
    \caption{Distributions of the ratio $R_\mathrm{en}$  between the total reconstructed energy $E_\mathrm{reco}^\mathrm{tot}$ and the total generated energy $E_\mathrm{gen}^\mathrm{tot}$. The results are obtained with the PFClustering algorithm, the DeepCluster model with a CNN-based center-finder CNN (CF CNN), and with a GNN-based center-finder (CF GNN). The distributions are obtained from the single-photon test dataset. A second peak $R_\mathrm{en}\approx 2$  arises for the CF CNN, while it is  eliminated with the CF GNN.}
  \label{figure:energy_sum_gnn}
\end{figure}

This effect is illustrated in Fig.~\ref{figure:energy_sum_gnn}, which presents the distribution of the ratio $R_\mathrm{en} = \Erec{\mathrm{tot}} / \Egen{\mathrm{tot}}$ where $\Erec{\mathrm{tot}}$ and $\Egen{\mathrm{tot}}$ are respectively is the total reconstructed energy $\Erec{\mathrm{tot}}$ (sum of the energies of all the reconstructed particles in the event) and the total generated energy $\Egen{\mathrm{tot}}$ (sum of the energies of all the generated particles in the event). The distribution of $R_\mathrm{en}$ is obtained from the single-photon dataset. The additional peak around two observed for the CNN-based center-finder (CF CNN on the figure) originates from the energy double counting.

The double counting issue is cured by changing the center-finder NN architecture as presented in the next section.


\subsection{Center-finder NN -- Graph Neural Network}
\label{section:cf_gnn}
The energy double counting is due to the fact that the CNN-based DeepCluster model does not receive information from the rest of the event, as a consequence, it is not aware of the existence of several seed windows corresponding to the same generated particle.

To solve the issue, we pass several neighboring seed windows as input to the network. This is achieved using a GNN architecture for the center-finder NN while the seed-finder NN remains unchanged. The GNN implementation provides in addition message-passing capabilities (\cite{zhou2021graph}) enabling information sharing between the different seed windows.


The inputs to the GNN-based center-finder are constructed as follows. First, the list of seed windows in the event is ordered based on the energy of its center crystal $E_\mathrm{seed}$. This list is processed as follows:
    \begin{enumerate}
        \item A center-finder input is initialized from the window $w_\mathrm{ref}$ with the highest energy $E_{seed}$ in the list. At this step, the input shape is $1\times7\times7$.
        \item For each remaining seed window $w_\mathrm{alt}$ in the list, the $\Delta R$ distance between $w_\mathrm{ref}$ and $w_\mathrm{alt}$ is computed. If $\Delta R < 3$, $w_{alt}$ is added to the input. After this step, the input shape is $N_\mathrm{w}\times7\times7$ where $N_\mathrm{w}$ is the number of added seed windows.
        \item All of the seed windows included in the input are removed from the list of seed windows. The process is iterated until this list is empty.
    \end{enumerate}

In this work, $N_\mathrm{w}$ is chosen to be at most 4. This maximum can be easily adjusted to higher values as well. For each input, if $N_\mathrm{w} < 4$, it is completed with $7\times7$ empty windows, \emph{i.e.} with null crystal energies. In such a way, the input shape is fixed to $4\times7\times7$.

The 4 seed windows of the input represent the nodes of the graph. In the center-finder NN, each seed window $j$ is first separately processed by a chain of convolutional layers (identical to the CNN center-finder implementation) in order to extract the vector of summary features $v_j$. The message-passing is implemented as the concatenation of these vectors. It results in 4 updated vectors $\bar{v}_j$, each of them containing information about their neighbors: 

\begin{equation}
\begin{aligned}
\bar{v}_1 &= \begin{pmatrix}
v_1 \\
v_2 \\
v_3 \\
v_4
\end{pmatrix},
&
\bar{v}_2 &= \begin{pmatrix}
v_2 \\
v_1 \\
v_3 \\
v_4
\end{pmatrix},
&
\bar{v}_3 &= \begin{pmatrix}
v_3 \\
v_1 \\
v_2 \\
v_4
\end{pmatrix},
&
\bar{v}_4 &= \begin{pmatrix}
v_4 \\
v_1 \\
v_2 \\
v_3
\end{pmatrix}
\end{aligned}
\end{equation}

\begin{figure}[t]
\centering
\includegraphics[width=\linewidth]{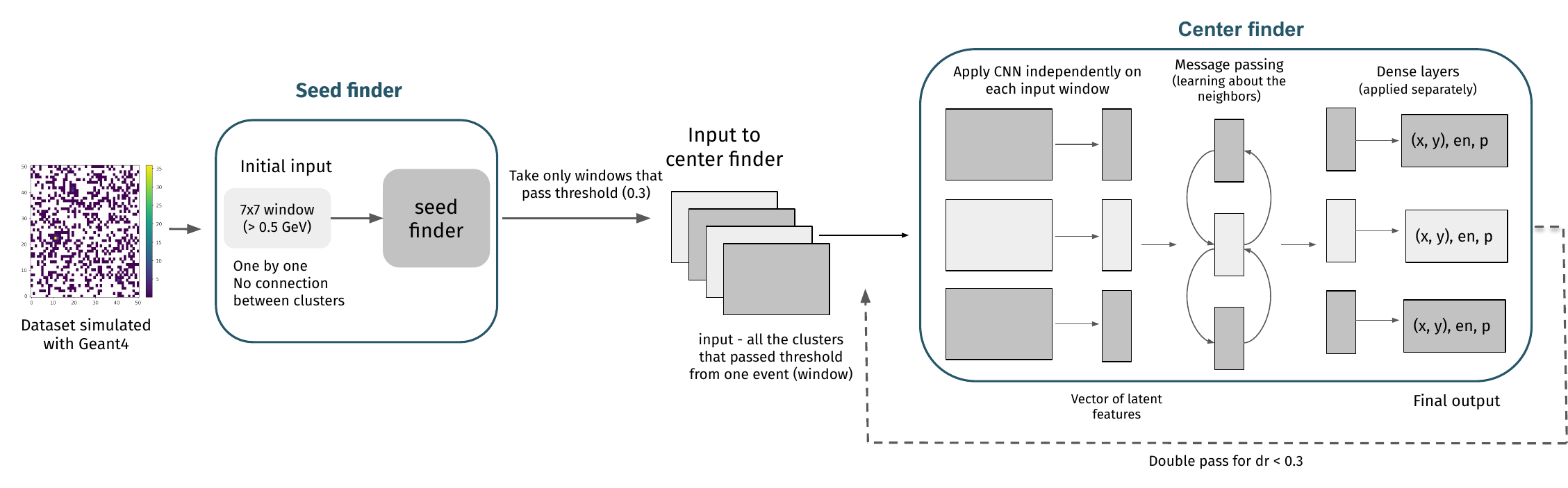} 
\caption{Flow chart of the DeepCluster model. $7\times7$ seed windows are first selected around all possible seeds ($>$0.5~GeV) in the event. They are separately passed as input to the seed-finder NN predicting $P_\mathrm{seedSF}$ for each seed window. Selected seed windows with $P_\mathrm{seedSF}$ $>$ $P_\mathrm{seedSF}^\mathrm{thr}$ are combined into groups of 4 with their neighbors and passed to center-finder NN which predicts the coordinates \xrec{}, \yrec{}, the energy \Erec{} and a new seed score $P_\mathrm{seedCF}$ for each seed window.}
\label{figure:network_pipeline}
\end{figure}

The combined vectors $\bar{v}_i$ are then passed independently to a set of dense layers until the final output is extracted. 
The GNN-based center-finder NN  predicts 4 values for each seed window $j$ in the input $i$: the kinematic variables (\xrec{ij}, \yrec{ij}), \Erec{ij} and a new seed score $P_\mathrm{seedCF}^{ij}$, indicating the likelihood to be associated to a true generated particle. 
In this version of the DeepCluster model, the seed-finder NN serves as an initial filter, separating signal from background, while the GNN-based center-finder corrects for the incorrect predictions related to double counting: eventually only objects with $P_\mathrm{seedCF} >P_\mathrm{seedCF}^\mathrm{thr}$ are selected. The optimization of $P_\mathrm{seedCF}^\mathrm{thr}$ is presented in Section~\ref{section:optimization}. The GNN-based center-finder architecture is presented in Fig.~\ref{figure:network_pipeline}. The loss used in the training is the sum of three terms related to the coordinates, the energy, and the seed probability predictions. The two first terms are based on a mean absolute error loss while the last one is based on a focal cross-entropy loss~\cite{focal_loss}. The combined loss is computed as:

\begin{equation} \label{eq:loss_function}
\begin{aligned}
\mathcal{L}_\mathrm{pos} & = \frac{1}{4\cdot N_b} \sum_{i=1}^{N_b}\sum_{j=1}^4  \frac{1}{2}
\left( \lvert \xrec{ij} - \xloc{ij} \rvert +  \lvert \yrec{ij}-\yloc{ij}\rvert \right) \\
\mathcal{L}_\mathrm{en } &= \frac{1}{4\cdot N_b} \sum_{i=1}^{N_b}\sum_{j=1}^4  \lvert \Erec{ij} - \Egen{ij}\rvert  \\
\mathcal{L}_\mathrm{seed} & = -\frac{1}{4\cdot N_b} \sum_{i=1}^{N_b}\sum_{j=1}^{4} \alpha(1-P^{ij}_\mathrm{seedCF})^\gamma \log(P^{ij}_\mathrm{seedCF}), \\
\mathcal{L}_\mathrm{tot} & = \mathcal{L}_\mathrm{pos} + \mathcal{L}_\mathrm{en} + k_\mathrm{s} \cdot \mathcal{L}_\mathrm{seed}, 
\end{aligned}
\end{equation}
where $N_b$ is the number of batches, the focal loss parameters $\gamma$ and $\alpha$ are chosen to be $\gamma=2$, $\alpha=0.25$,  and $k_\mathrm{s}$ is an adjustable parameter associated with the seed loss, it is referred to as \textit{seed-loss weight}.

\subsubsection{Results}
In the GNN implementation of the center-finder NN, the model receives information about all the neighboring seed windows simultaneously. With this adjustment, the network is able to make a more informed decision for each seed window and additionally better attribute energy fractions for different windows. 

The related improvement is shown in Fig.~\ref{figure:energy_sum_gnn} where the distribution of $R_\mathrm{en}$ obtained with the single-photon dataset is presented. For the center-finder GNN (CF GNN) only the seed windows with $P_\mathrm{seedCF} > $ 0.4 are selected. One can notice the disappearance of the peak at 2 signalling the resolution of the double-counting issue.

\section{Network optimization}\label{section:optimization}
The DeepCluster model has a number of adjustable parameters. This section describes the optimization of these parameters to achieve the best possible performance. This starts with the optimization of the seed loss weight, followed by the optimization of the different seed-probability thresholds: $P_\mathrm{seedSF}^\mathrm{thr}$ and $P_\mathrm{seedCF}^\mathrm{thr}$. Then another step is added to the algorithm in order to suppress multiple predicted particles arising from the same generated one. Eventually, the adjustment of the hyperparameters which underlie the different NNs is presented.

\subsection{Seed-loss weight}
The GNN-based center finder contains in its loss a term related to the predicted seed probability which is weighted by the parameter $k_\mathrm{s}$. The definition of the total loss is given by Eq.~\ref{eq:loss_function}, where the parameter $k_\mathrm{s}$ is introduced.
The optimal value of $k_\mathrm{s}$ is obtained by comparing the performance achieved with different seed-loss weights $k_\mathrm{s}$. The evolution of the losses for the training and validation datasets are shown in \nameref{appendix_d} for different values of $k_s$.  

\subsection{Seed-probability thresholds}
In the final DeepCluster model implementation, the seed-finder NN and the center-finder NN predict two different seed scores: $P_\mathrm{seedSF}$, $P_\mathrm{seedCF}$. Solely the seed windows fulfilling the criteria $P_\mathrm{seedSF}>P_\mathrm{seedSF}^\mathrm{thr}$ and $P_\mathrm{seedCF}>P_\mathrm{seedCF}^\mathrm{thr}$ are kept for further analysis, the other ones are discarded as background windows.
The energy and position resolutions as well as the signal efficiencies and background rates are studied on the single-photon test sample for different sets of thresholds. We retain the values $P_\mathrm{seedSF}^\mathrm{thr}=0.3$ and $P_\mathrm{seedCF}^\mathrm{thr}=0.4$ as they ensure excellent performance in terms of position and energy resolutions ($\sigma_x\sim0.02$~crystal and $\sigma_E\sim0.56~\GeV$) while maintaining a high signal efficiency ($\epsilon\sim99.5~\%$).


\subsection{Particle splitting}
One of the advantages of the DeepCluster model resides in its ability to reconstruct close-by particles. Still, for both the DeepCluster and the PFClustering algorithms, the closer the particles are, the harder it is to disentangle them in the reconstruction process. There is a limit to the distance between two particles below which the reconstruction is not reliable.

\begin{figure}[ht]
\centering
\includegraphics[width=0.5\linewidth]{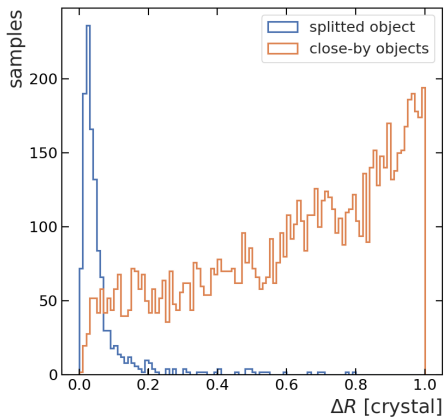} 
\caption{Distributions of $\Delta R$ between two close-by generated particles associated to two reconstructed objects (labeled generated particle), obtained from the two-photon dataset, and $\Delta R$ between two close-by reconstructed objects arising from the same particle in the single-photon dataset (labeled reco-object).}
\label{figure:separation}
\end{figure}

This limit is explored in Fig.~\ref{figure:separation} by comparing the minimal distance between two properly reconstructed objects (\emph{i.e.} associated to two generated particles) to the minimal distance between two reconstructed objects associated to the same particle. The latter case corresponds to a generated particle giving rise to two very close-by clusters, thus splitting in parts the energy of the original particle, this is referred to as particle splitting. From this figure, one can see that for $\Delta R <0.3$, the reconstruction of two close-by particles solely corresponds to the splitting of a single particle. On the opposite, for $\Delta R>0.3$, close-by particles are properly reconstructed and single-particle splitting is nonexistent.


In order to mitigate single-particle splitting while maintaining high signal efficiency for two close-by photons, a dedicated procedure is implemented. We first group reconstructed objects with $\Delta R <0.3$, as aforementioned these group contains essentially particle-splitting clusters. In these groups, the highest energy object is kept while the others are discarded. The seed window associated to the corresponding selected object is passed a second time in the center-finder NN, for this second pass, the neighboring seed windows associated to the discarded objects are not considered. This second pass allows to keep a single reconstructed object for a single particle while properly predicting its energy thus avoiding the particle-splitting phenomena. 


\subsection{Hyperparameters}
The tuning of the hyperparameters of the center-finder network is performed using a Bayesian optimization~\cite{keras_optimization}. The optimal parameters are presented in Tab.~\ref{table:hyperparameter}. 
The final DeepCluster model is trained using the LAMB optimizer~\cite{lamb}. The epoch achieving the lowest value in the validation dataset is selected.

\begin{table}[ht!]
  \centering
  \renewcommand{\arraystretch}{1.5}
  \setlength{\tabcolsep}{10pt}
  \begin{tabular}{p{4cm}|p{7cm}}
    \toprule
    \textbf{Parameter} & \textbf{Value} \\
    \midrule
    1$^\mathrm{st}$ convolutional layer & number of filters: 128, kernel size: 3 + batch normalization \\
    2$^\mathrm{nd}$ convolutional layer & number of filters: 112, kernel size: 1 + batch normalization \\
    Common dense layers & nodes: 1100, dropout: 0.1 \\
    Center dense layer & nodes: 500, dropout: 0.3 \\
    Energy dense layer & nodes: 100, dropout: 0.1 \\
    Seed dense layer & nodes: 250, dropout: 0.3 \\
    Batch size & 512 \\
    Learning rate & 0.0001 \\
    \bottomrule
  \end{tabular}
  \caption{Final values for the DeepCluster model after the Bayesian hyperparameter optimization was performed for center-finder NN.}
  \label{table:hyperparameter}
\end{table}

\section{DeepCluster performance}\label{section:results}
This section presents the performance of the optimized DeepCluster model and compares it to the one obtained with the PFClustering algorithm. The energy and position resolutions are evaluated as half the interval containing 68\% of the distribution and centered on the median. In addition to the position and energy resolutions, other important metrics are investigated: signal efficiency, background yield, and particle-splitting yield, as defined in Table~\ref{table:seed_finder_evaluation}.
\begin{table}[ht]
  \centering
  \renewcommand{\arraystretch}{1.5}
  \setlength{\tabcolsep}{10pt}
  \begin{tabular}{l|p{0.7\linewidth}}
    \toprule
    \textbf{Name} & \textbf{Description} \\
    \midrule
    Signal efficiency & Number of correctly reconstructed objects divided by the number of generated particles. \\
    Particle-splitting yield & Number of events where one particle is reconstructed as two different objects, reported for 100k photons. \\
    Background yield   & Number of reconstructed objects that do not correspond to any generated particle, reported for 100k toy simulations (51 crystals $\times$ 51 crystals). \\
    \bottomrule
  \end{tabular}
  \caption{Description of the variables used for evaluation of the algorithms.}
  \label{table:seed_finder_evaluation}
\end{table}
The matching procedure linking the reconstructed objects with the true generated particles to determine if reconstructed objects are to be considered as signal or background is described in \nameref{appendix_c}. 
Each of the reconstructed objects is linked to only one generated particle, while the generated particle can be linked to multiple reconstructed objects. In the latter case, the objects are tagged as particle-splitting. The same matching procedure is applied for the objects reconstructed by the DeepCluster model and PFClustering algorithm.

First, the DeepCluster model is tested with photon datasets that are comparable to the ones used for training. In a second step, the DeepCluster reconstruction is tested with electrons and finally, the algorithm is applied to the reconstruction of $\pi^0$-meson decays producing collimated photons in the detector.

\subsection{Performance for photons}
The performance for single-photon and two-photon test datasets is presented in this section. 

\begin{figure}[ht]
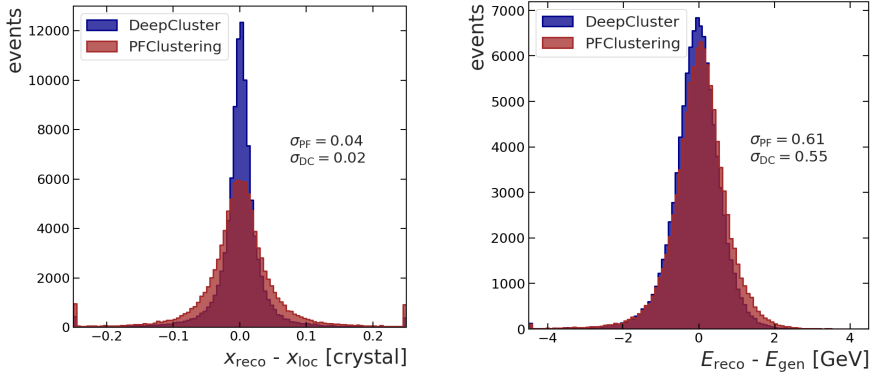

  \begin{minipage}{0.46\textwidth}
    \centering
    \includegraphics[width=\textwidth]{figures/x_distribution.pdf}
  \end{minipage}\hspace{0.5cm}
  \begin{minipage}{0.46\textwidth}
    \centering
        \includegraphics[width=\textwidth]{figures/en_distribution.pdf}
  \end{minipage}
  \caption{Left: distribution of the difference between the reconstructed position $x_\mathrm{reco}$ and the generated position of the particle $x_\mathrm{loc}$. Right: distribution of the difference between the reconstructed energy $E_\mathrm{reco}$ and the generated energy of the particle $E_\mathrm{gen}$. The results are obtained by applying the PFClustering algorithm and DeepCluster model on the single-photon test dataset.}
  \label{fig:resolutions}
\end{figure}

Figure~\ref{fig:resolutions} shows the distributions of the difference between the reconstructed and generated values for position (left) and energy (right) in the single-photon dataset.
The position resolution is improved by 50\% with regard to the PFClustering algorithm while the energy resolution is improved by about 10\%. The improvement is more drastic for the two-photon dataset, where both the position and energy resolutions are improved by about 60\%. 

\begin{figure}[ht]
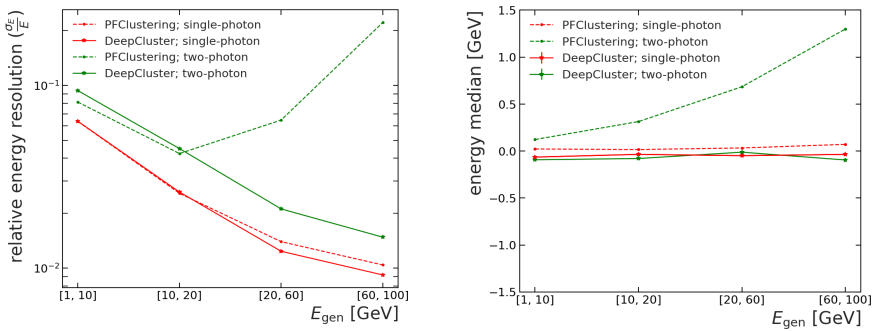

  \begin{minipage}{0.46\textwidth}
    \centering
    \includegraphics[width=\textwidth]{figures/energy_resolution_1.pdf}
  \end{minipage}\hspace{0.5cm}
  \begin{minipage}{0.46\textwidth}
    \centering
        \includegraphics[width=\textwidth]{figures/energy_median.pdf}
  \end{minipage}
  \caption{Relative energy resolution (left) and energy median (right) obtained with the DeepCluster model and PFClustering algorithm applied on the single- and two-photon test datasets. The results are shown in the bins of generated energy $E_\mathrm{gen}$.}
  \label{fig:energy_resolution}
\end{figure}

\begin{figure}[ht]
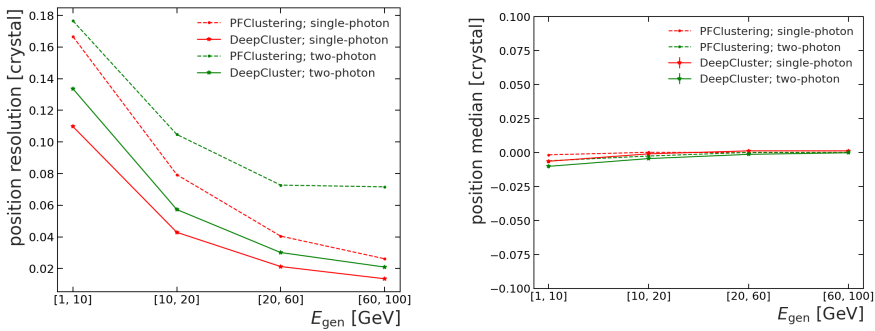

  \begin{minipage}{0.46\textwidth}
    \centering
    \includegraphics[width=\textwidth]{figures/position_resolution.pdf}
  \end{minipage}\hspace{0.5cm}
  \begin{minipage}{0.46\textwidth}
    \centering
        \includegraphics[width=\textwidth]{figures/position_median.pdf}
  \end{minipage}
  \caption{Position resolution (left) and position median (right) obtained with the DeepCluster model and PFClustering algorithm applied on the single- and two-photon test datasets. The results are shown in the bins of generated energy $E_\mathrm{gen}$.}
  \label{fig:position_resolution}
\end{figure}

This is illustrated in Figs.~\ref{fig:energy_resolution} and ~\ref{fig:position_resolution} where the standard deviation (left) and median (right) of the distributions of the difference between the reconstructed and generated values for energy and position are presented as a function of the generated energy of the particle $E_\mathrm{gen}$, for the single- and two-photon datasets. For the energy, the relative resolution is shown. In the two-photon case, the resolutions, for energies above 20 \GeV, are improved by more than 70 \% for the energy and 60\% for the position. The bias observed for the median of the distributions for the DeepCluster model is small compared to the  corresponding resolution. It can be either fixed by further optimizing the parameters of the network or corrected for.   

\begin{figure}[ht]
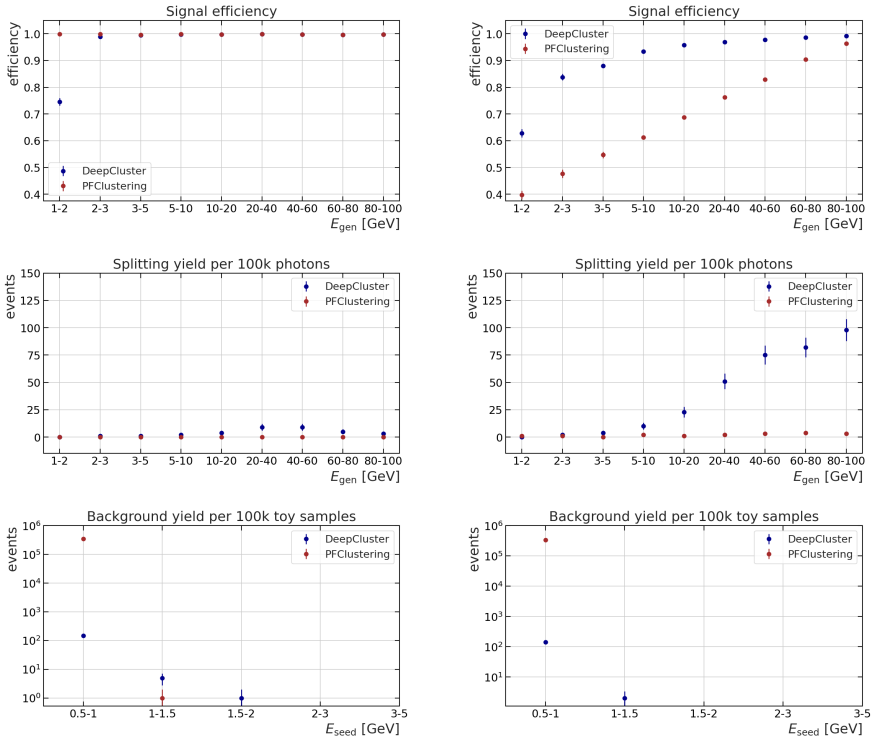

  \begin{minipage}{0.46\textwidth}
    \centering
    \includegraphics[width=\textwidth]{figures/final_photon_signal-1.pdf}
  \end{minipage}\hspace{0.5cm}
  \begin{minipage}{0.46\textwidth}
    \centering
        \includegraphics[width=\textwidth]{figures/final_photon_signal_2-1.pdf}
  \end{minipage}
  \caption{Signal efficiency (top), splitting yield for 100k photons (middle), and background yield for 100k toy simulations (bottom) values. The results are obtained by applying the PFClustering and DeepCluster model on the single-photon (left) and two-photon (right) test datasets.}
  \label{fig:signal_final_results}
\end{figure}

Figure~\ref{fig:signal_final_results} presents the signal efficiency, particle-splitting yield, and background yield for the DeepCluster model and the PFClustering algorithm for the single-photon dataset on the left and for the two-photon dataset on the right. The results are presented as a function of the energy of the generated particle $E_\mathrm{gen}$ for the signal efficiency and splitting yield and as a function of the energy of the seed crystal $E_\mathrm{seed}$ for the background yield. In the single-photon case, the signal efficiency is the same as for PFClustering starting from about 5 \GeV, while the background rate, coming from noise reconstructed as low energy clusters, is improved by a factor of about 2000. In the two-photon case, the signal efficiency is largely improved, up to a factor two at low energy.

A summary of the performance is presented in Table~\ref{tab:summary_result}. The DeepCluster model outperforms the PFClustering algorithm in terms of position and energy resolution both for the single- and two-photon cases .
Most notably, the signal efficiency for the two-photon dataset obtained with the DeepCluster model is 97.0$\%$, while with PFClustering, it is only 82.0$\%$.

\begin{table}[htbp]
  \centering
  \renewcommand{\arraystretch}{1.5}
  \begin{tabular}{ccccc}
    \toprule
     & \multicolumn{2}{c}{Single-Photon} & \multicolumn{2}{c}{Two-Photon} \\
    \cmidrule(lr){2-3} \cmidrule(lr){4-5}
    & PFClustering & DeepCluster & PFClustering & DeepCluster \\
    \midrule
    $\sigma_x$~[crystal]  & 0.04 $\pm$ 0.00 & 0.02 $\pm$ 0.00 & 0.08 $\pm$ 0.00 & 0.03 $\pm$ 0.00 \\
    $\sigma_E$~[GeV] & 0.61 $\pm$ 0.00 & 0.55 $\pm$ 0.00 & 6.24 $\pm$ 0.01 & 0.92 $\pm$ 0.00 \\
    \midrule
    $\epsilon$ & 99.8 $\pm$ 0.3 \% & 99.5 $\pm$ 0.3\% & 82.0 $\pm$ 0.3\% & 97.0 $\pm$ 0.3\% \\
    $N_\mathrm{split}/100\mathrm{k~photons}$ & 0 & 34 $\pm$ 6 & 17 $\pm$ 4 & 345 $\pm$ 19 \\
    $N_\mathrm{bkg}/100\mathrm{k~toy~samples}$ & 350k $\pm$ 19k & 153 $\pm$ 12 & 320k $\pm$ 18k & 146 $\pm$ 12 \\
    \bottomrule
  \end{tabular}
  \caption{Performance comparison for position and energy resolutions, signal efficiency, splitting yield for 100k photons, and background yield for 100k toy simulation between PFClustering and DeepCluster algorithms for single- and two-photon datasets.}
  \label{tab:summary_result}
\end{table}

\newpage
\subsection{Electrons}

The DeepCluster algorithm is tested on the electron dataset. Although the network is not specifically trained on it, it still shows excellent performance. The resolutions in energy and position, the signal efficiency, and the splitting and background yields are extremely similar to the ones obtained for photons, as expected in absence of magnetic field and material in front of the calorimeter.



\newpage
\subsection{Neutral pions}
Finally, the results achieved on the $\pi_0$ sample are presented. In this case, the reconstruction algorithms have to detect both of the photons originating from the $\pi_0$ decay and correctly estimate their energy. With this information, the mass of the $\pi_0$ can be reconstructed as: 

\begin{equation}
    m_{\gamma \gamma} = \sqrt{2E^1_\mathrm{reco} E^2_\mathrm{reco} \left(1-\cos {\theta_\mathrm{reco}}\right)}, 
\end{equation}
where $E^1_\mathrm{reco}$, $E^2_\mathrm{reco}$ are the reconstructed energies of two photons and $\theta_\mathrm{reco}$ is the reconstructed angle between them.

\begin{figure}[h!]
\centering
\includegraphics[width=\linewidth]{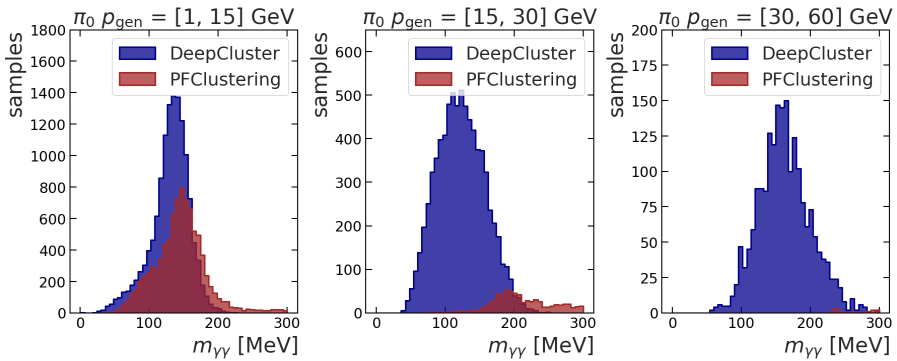} 
\caption{$m_{\gamma \gamma}$ mass distribution reconstructed with DeepCluster model and PFClustering algorithm on the $\pi_0$ dataset. The results are shown in bins of the generated momentum $p_\mathrm{gen}$ of the $\pi_{0}$.}
\label{fig:pion}
\end{figure}

The diphoton mass distributions reconstructed with the DeepCluster model and PFClustering are presented in Fig.~\ref{fig:pion}. The results are shown in bins of the generated momentum $p_\mathrm{gen}$ of the $\pi_{0}$. The model achieves excellent results, outperforming the PFClustering $\pi_0$ detection efficiency by a factor of more than two. Moreover, the diphoton mass resolution is significantly better with the model. 

As in the case of electrons, the model is not specifically trained on the $\pi_0$ dataset. Moreover, the photons enter the toy calorimeter under different angles and not perpendicularly as in the training sample. This further underscores the robustness of the network. 

\section{Conclusion}
This paper introduces an innovative machine-learning algorithm, called the DeepCluster model, to measure the energy and position of photons and electrons based on convolutional and graph neural networks, taking the geometry of the CMS electromagnetic calorimeter as an example. 


To develop the DeepCluster model and evaluate its performance, a dedicated simplified simulation of the ECAL is created, and different implementations of the model are tested. A two-step network strategy, incorporating both CNN and GNN architectures, delivers the most optimal performance and effectively addresses all identified issues.
The final model is tested on datasets with single photons
and two overlapping photons, as well as on datasets with electrons or
neutral pions. In all cases, the DeepCluster shows superior performance compared to the method currently in use in CMS in terms of coordinate and energy resolutions, as well as background rejection and signal efficiency.
In particular, the DeepCluster model demonstrates excellent results in distinguishing
between closely spaced particles, reconstructing approximately twice
as many $\pi^0$ as the traditional approach. 

These results demonstrate that this approach is very promising to enhance the performance of calorimeters in high-energy physics experiments.
Moreover, as the network processes only small 7$\times$7 windows, the scalability for the full ECAL will not pose a problem. The particles being either far apart in the calorimeter (single-photon dataset) or in close proximity (two-photon dataset), all physics cases are covered.
Finally, the performance gain of this approach could be even larger in presence of pile-up interactions, as the local information could allow  the network to better mitigate their impact.
\section*{Acknowledgments}
This project has received funding from the European Union's Horizon 2020 research and innovation programme under grant agreement No 800945 — NUMERICS — H2020-MSCA-COFUND-2017.
 We gratefully thank the CMS experiment from which this work was inspired.
 We also acknowledge the CEA (France) for financial and computing support and Dr. James Rich and Dr. Nathalie Besson for their careful reading of this manuscript.

\section*{Appendix A}\label{appendix_a}

The BDT is trained on a single-photon dataset with a flat energy distribution between 1 and 250~GeV. The hyperparameters of the model, such as \textit{learning rate}, \textit{number of estimators}, \textit{minimum split}, \textit{minimum leaf}, and \textit{maximum depth}, are optimized to achieve the best performance using a random grid search.  

The resulting regression scores for 20 different sets of parameters are calculated. The full summary of optimization is presented in Table~\ref{table:bdt_optimization} and parameters corresponding to the highest score (trial 10) are selected based on these results.

\begin{table}
\centering
\renewcommand{\arraystretch}{1.5}
\setlength{\tabcolsep}{8pt}
\begin{tabular}{l | p{0.1\linewidth} p{0.15\linewidth}  p{0.1\linewidth} p{0.1\linewidth} p{0.1\linewidth} p{0.1\linewidth}}
\toprule
\textbf{trial} & \textbf{learning rate}& \textbf{number of estimators} & \textbf{minimum split} & \textbf{minimum leaf} & \textbf{maximum depth} & \textbf{score} \\
\midrule
 1 & 0.20 & 100 & 20 & 5 & 20 & 0.7799 \\
 2 & 0.05 & 300 & 10 & 10 & 2 & 0.7676 \\
 3 & 0.20 & 300 & 10 & 5 & 10 & 0.7815 \\
 4 & 0.05 & 300 & 20 & 2 & 2 & 0.7756 \\
 5 & 0.20 & 300 & 2 & 10 & 5 & 0.7886 \\
 6 & 0.05 & 500 & 4 & 5 & 20 & 0.7797 \\
 7 & 0.20 & 100 & 10 & 2 & 10 & 0.7952 \\
 8 & 0.05 & 100 & 20 & 2 & 5 & 0.7874 \\
 9 & 0.20 & 500 & 20 & 2 & 5 & 0.7947 \\
 10 & 0.10 & 100 & 2 & 2 & 10 & 0.7953 \\
 11 & 0.20 & 300 & 10 & 2 & 10 & 0.7887 \\
 12 & 0.10 & 300 & 2 & 10 & 2 & 0.7754 \\
 13 & 0.10 & 300 & 4 & 10 & 10 & 0.7850 \\
 14 & 0.20 & 300 & 2 & 10 & 10 & 0.7782 \\
 15 & 0.20 & 500 & 20 & 2 & 2 & 0.7913 \\
 16 & 0.05 & 100 & 2 & 5 & 10 & 0.7913 \\
 17 & 0.10 & 300 & 4 & 5 & 5 & 0.7951 \\
 18 & 0.10 & 500 & 20 & 10 & 10 & 0.7828 \\
 19 & 0.20 & 300 & 4 & 10 & 10 & 0.7782 \\
 20 & 0.05 & 300 & 2 & 2 & 5 & 0.7952 \\
\bottomrule
\end{tabular}
\caption{Results for the BDT hyperparameter optimization. The algorithm is applied on the single-photon validation dataset, and the regression score measures its performance. The parameters of the 10$^\mathrm{th}$ trial are chosen as optimal as they correspond to the highest value of the regression score.}
\label{table:bdt_optimization}
\end{table}

\begin{figure}[ht]
\centering
\includegraphics[width=0.85\linewidth]{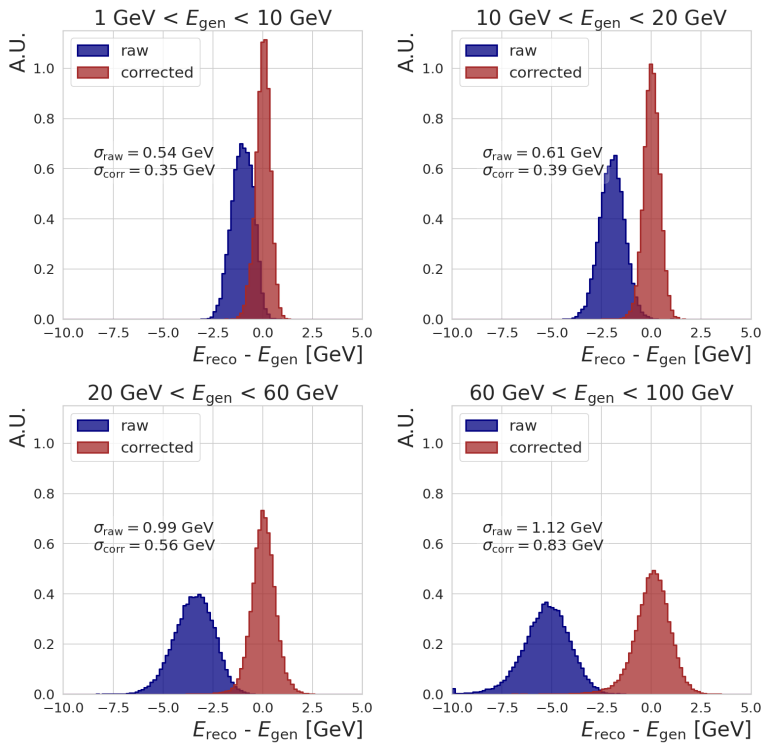} 
\caption{Distribution of the difference between the reconstructed energy $E_\mathrm{reco}$ and the generated energy of the particle $E_\mathrm{gen}$. The results are obtained by applying the PFClustering algorithm (raw) and the PFClustering algorithm with additional energy regression (corrected) on the single-photon test dataset. The distributions are presented in four bins of the generated photon energy $E_\mathrm{gen}$: [1, 10]~GeV, [10, 20]~GeV, [20, 60]~GeV, and [60, 100]~GeV. The resolutions evaluated from these distributions are reported on the plots.}
\label{figure:pfclustering_bdt}
\end{figure}

The performance in terms of energy for the final optimized model is shown in Fig.~\ref{figure:pfclustering_bdt} along with the comparison to the raw PFClustering prediction. With the BDT correction, the energy resolution is drastically improved compared to the raw results: by 35$\%$, 36$\%$, 43$\%$, and 26$\%$ for the energies of generated particles $E_\mathrm{gen}$~=~[1, 10]~GeV, $E_\mathrm{gen}$~=~[10, 20]~GeV, $E_\mathrm{gen}$~=~[20, 60]~GeV, and $E_\mathrm{gen}$~=~[60, 100]~GeV, respectively. Moreover, the corrected energy distributions are centered around zero, unlike the ones obtained only from the raw PFClustering algorithm. 

\section*{Appendix B}\label{appendix_b}
\begin{figure}[ht]
  \begin{minipage}{0.49\textwidth}
    \centering
    \includegraphics[width=\textwidth]{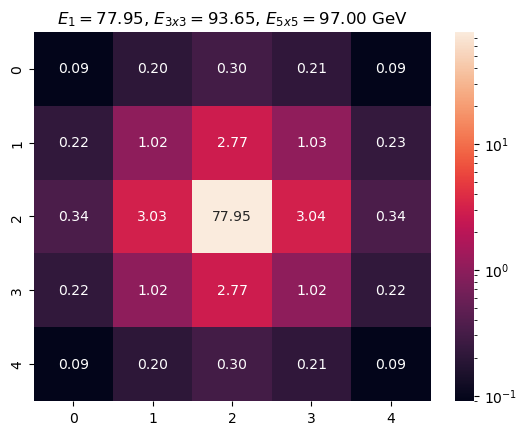}
  \end{minipage}
  \begin{minipage}{0.49\textwidth}
    \centering
        \includegraphics[width=\textwidth]{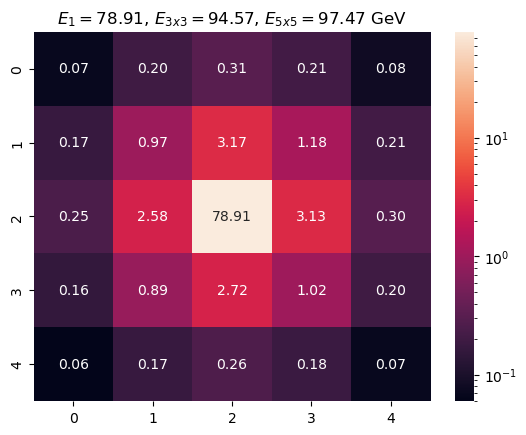}
  \end{minipage}
  \caption{Energy deposit profiles from the toy calorimeter (left) and from Geant4 simulation of ECAL (right)~\cite{ecal_beam}. The profiles are obtained with an electron beam.}
\label{fig:energy_deposit_simulation}
\end{figure}

To validate the simulation of the toy calorimeter, the energy deposit profile is investigated and compared to the simulation of the actual ECAL. In order to do it, 1,000 electrons at 100~GeV are shot at the center of the detector, and the average deposited energy in each cell of the 5 x 5 crystal matrix around the central crystal is calculated. The results are shown in Fig.~\ref{fig:energy_deposit_simulation} (left) and they are compared to the ones achieved with the real ECAL simulation presented in Fig.~\ref{fig:energy_deposit_simulation} (right).

The obtained results are comparable: the average ratio of energy deposits of initial electrons in the central crystal (E1), $3\times3$ matrix (E3), and $5\times5$ (E5) matrix around it are approximately 78$\%$, 94$\%$, 97$\%$ for toy calorimeter and 79$\%$, 95$\%$, 98$\%$ for ECAL simulation~\cite{ecal_beam}, respectively. This demonstrates that the toy calorimeter can be used as a proxy for the real ECAL. 

The asymmetry in the energy deposition around the central crystal is not present for the toy calorimeter (unlike in the ECAL simulation) as it does not include crystal tilt.

\section*{Appendix C}\label{appendix_c}

A matching procedure that links the reconstructed objects with the true generated particles is applied to determine if reconstructed objects are to be considered as signal or background. A “matching” variable $r_\mathrm{match}$ is defined:
    \begin{equation}
    r_{\mathrm{match}} = \sqrt{\left(\frac{R_{\mathrm{reco}} - R_{\mathrm{loc}}}{\overline{R}}\right)^2 + \left(\frac{E_{\mathrm{reco}} - E_{\mathrm{gen}}}{\overline{E}}\right)^2 }, 
    \end{equation}
    where $R_\mathrm{\{reco,loc\}} = \sqrt{x_\mathrm{\{reco,loc\}}^2 + y_\mathrm{\{reco,loc\}}^2}$, $\overline{R}$ is the mean value of the ($R_{\mathrm{reco}}$ - $R_{\mathrm{loc}}$) distribution and $\overline{E}$ is the mean value of the ($E_{\mathrm{reco}}$ - $E_{\mathrm{gen}}$) distribution. These values are obtained from a preliminary truth association based only on the position. For each reconstructed object in the sample:
    \begin{itemize}
        \item $r_{\mathrm{match}}$ is computed for the considered reconstructed object and all the generated particles;
        \item A link is created between the reconstructed object and the particle with the smallest $r_{\mathrm{match}}$;
        \item If $\Delta R~=~\sqrt{(x_\mathrm{reco} - x_\mathrm{loc})^2 + (y_\mathrm{reco} - y_\mathrm{loc})^2}$~$>$~1.5~crystal, the link is removed and the reconstructed object is considered as background.
    \end{itemize}
    
\section*{Appendix D}\label{appendix_d}
\begin{figure}[h]
    \includegraphics[width=0.95\linewidth]{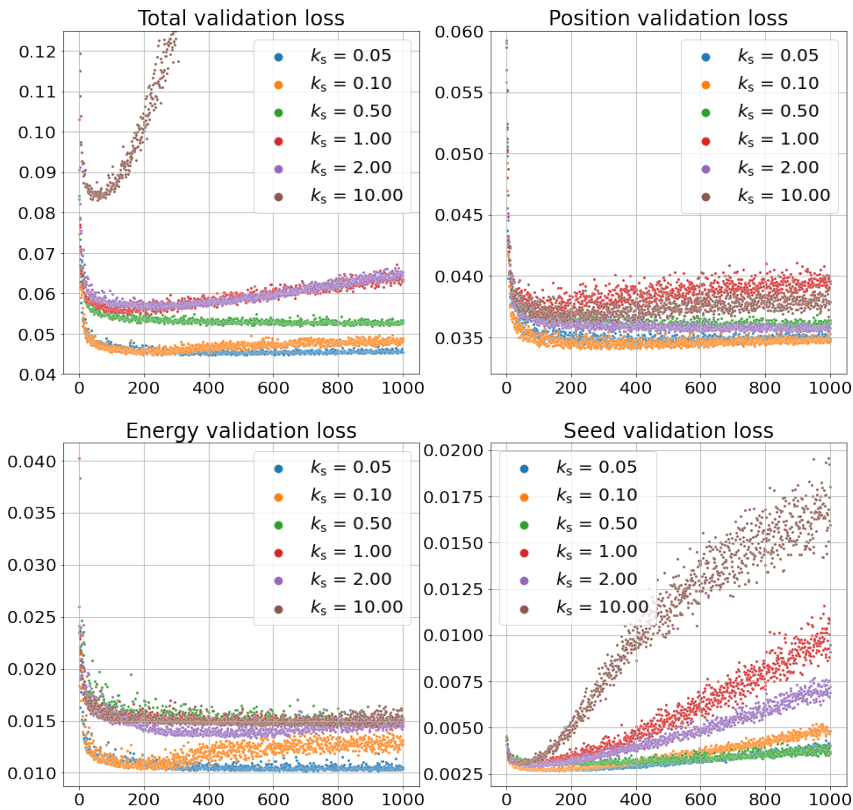} 
    \caption{The validation loss functions evolution versus epoch, presented for different seed weights ($k_\mathrm{s}$). The position, energy, seed, and total validation losses are shown.}
  \label{figure:loss_plots}
\end{figure}

\begin{figure}[ht]
\centering
    \includegraphics[width=0.60\linewidth]{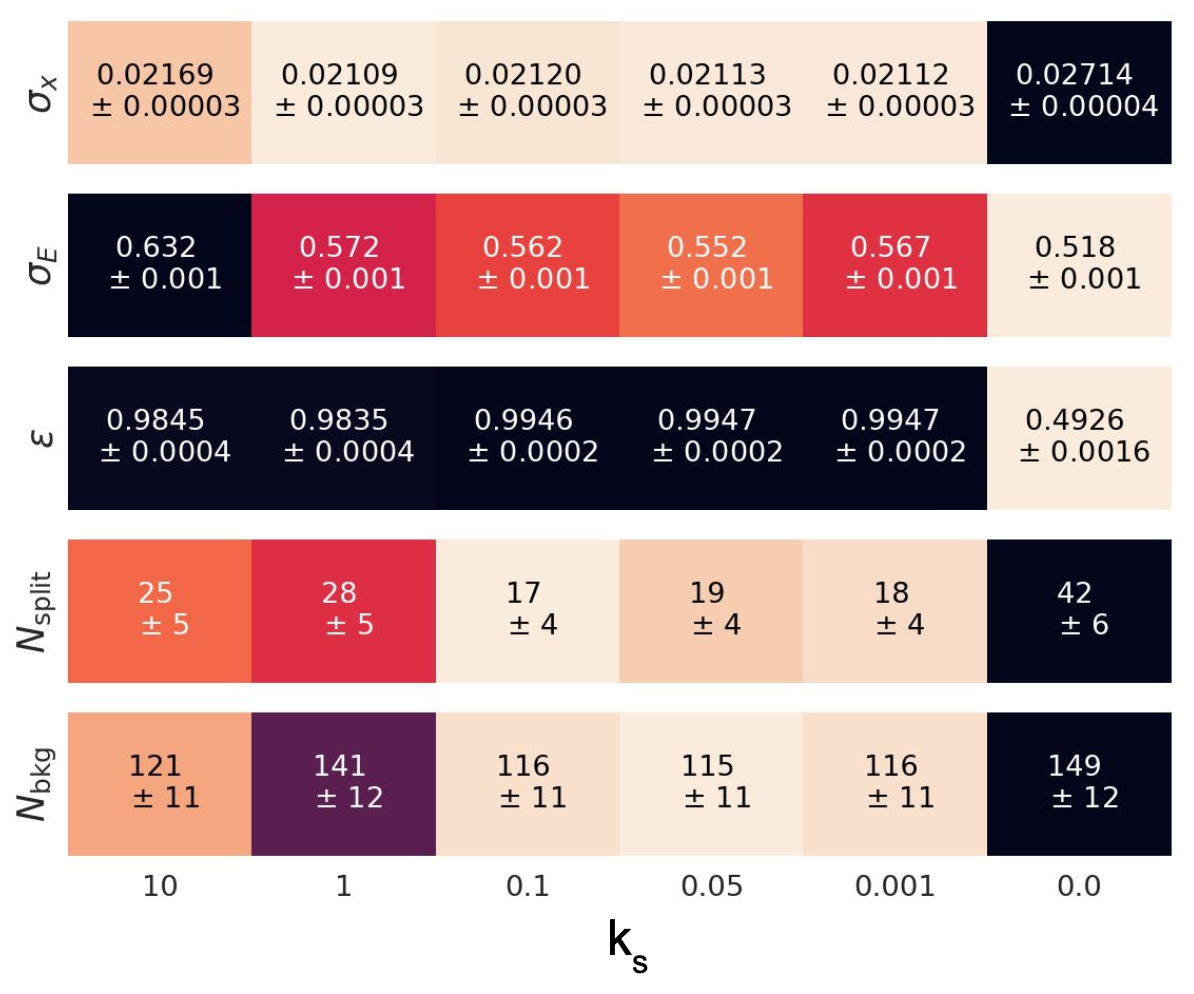} 
    \caption{Values corresponding to different physics performance metrics, as detailed in Section~\ref{section:results}, are obtained from scanning various $k_\mathrm{s}$ penalty terms. $k_\mathrm{s}=0.05$ is found to be the value yielding the best performance.}
  \label{figure:k_s_scan}
\end{figure}

The three terms in the loss function referenced in Eq.~\ref{eq:loss_function} collectively contribute to the minimization of the total loss function. The part of the neural network responsible for predicting the seed probability has shown a tendency for overfitting. This observation is supported by the divergence of the validation sample loss, as depicted in Fig.~\ref{figure:loss_plots}.
Predominantly, a higher contribution from the seed probability loss dominates the total validation loss. This behavior results in two undesirable effects. First, the total loss begins to diverge on the validation sample, causing the epoch corresponding to the minimum validation total loss to represent a less-than-ideal configuration for the network's other two outputs. Secondly, the gradients are disproportionately affected by the seed loss, leading to a suboptimal training for the remaining outputs. To mitigate these effects, we introduced a penalty term into the seed probability loss function. Based on the physics performance metrics associated with different penalty terms shown in Fig.~\ref{figure:k_s_scan}, we determined that applying a penalty term of $k_\mathrm{s}=0.05$ to the seed probability loss produces a reduced cumulative loss across all three networks. The model configuration at the epoch $\mathrm{n} = 321$ is considered as the most optimal since it has the smallest validation loss value and all of the three terms of the loss function are converged. 

\bibliography{sn-bibliography.bib}



\end{document}